\documentclass[12pt,dvips]{article}
\usepackage{epsfig}

\vbox{
\hbox{\hskip4.9in KSUCNR-114-00}
\hbox{\hskip4.9in ADP-00-47/T428}
\hbox{\hskip4.9in September 2000}
}
\def\lskip{\vspace{0.5cm}}

\def\bi{\begin{itemize}}
\def\ei{\end{itemize}}
\def\bc{\begin{center}}
\def\ec{\end{center}}
\def\bdm{\begin{displaymath}}
\def\edm{\end{displaymath}}
\begin{document}
\bc
{\bf \large  Measurement of the ${ \mathbf {F_2^n/F_2^p} }$ and 
${ \mathbf {d/u} }$ Ratios  in Deep Inelastic Electron Scattering off
$^3$H and $^3$He}
\footnote{Work supported in part by the U.S. Department of Energy and 
National Science Foundation, and the Australian Research Council.}
\ec
\vskip0.35cm
\bc
G.~G. Petratos$^1$, I.~R.~Afnan$^2$, F.~Bissey$^{3,4}$, J.~Gomez$^5$, 
A.~T.~Katramatou$^1$, W.~Melnitchouk$^{3,5}$ and A.~W.~Thomas$^3$
\vskip0.2in
{\it $^1$Kent State University, Kent, Ohio 44242, USA}\\
{\it $^2$Flinders University of South Australia, Bedford Park 5042,
  Australia}\\
{\it $^3$University of Adelaide, Adelaide 5005, Australia}\\
{\it $^4$Universit\'{e} Blaise Pascal, 63177 Aubi\`{e}re Cedex, France}\\
{\it $^5$Jefferson Lab, Newport News, Virginia 23606, USA}\\

\lskip
\lskip
ABSTRACT\\
\ec
\begin{quote}
\noindent 
\small 
We discuss a possible measurement of the ratio of the nucleon structure
functions, $F_2^n/F_2^p$, and the ratio of up to down quark
distributions, $u/d$, at large $x$, by performing deep inelastic 
electron scattering from the $^3$H and $^3$He mirror nuclei
with the 11 GeV upgraded beam of Jefferson Lab.
The measurement is expected to be almost free of nuclear effects,
which introduce a significant uncertainty in the extraction of these
two ratios from deep inelastic scattering off the proton and deuteron.
The results are expected to test 
perturbative and non-perturbative mechanisms of spin-flavor symmetry
breaking in the nucleon, and constrain
the structure function parametrizations needed for the
interpretation of high energy $e$-$p$, $p$-$p$ and $p$-$\bar{p}$
collider data.
The precision of the expected data can also allow for testing
competing parametrizations of the
nuclear EMC effect and provide valuable constraints on models of its
dynamical origin.
\end{quote}

\lskip
\lskip

\centerline{Invited talk presented at the International Workshop on the}
\centerline{\it Nucleon Structure in High x-Bjorken Region (HiX2000)}
\centerline{Temple University, Philadelphia, PA, USA}
\centerline{March 30 - April 1, 2000}

\newpage

\section{Introduction}

Measurements of the proton and deuteron structure functions have 
been of fundamental importance in establishing the internal quark structure
of the nucleon~\cite{ta91}.  The first evidence for the
presence of point-like constituents (partons) in the nucleon came from the
observation that the ratio of inelastic to Mott cross sections, measured
in the pioneering SLAC experiments, exhibited only small variation with
momentum transfer~\cite{bl69}.
The subsequent detailed analysis of the SLAC data~\cite{fr72} revealed
the predicted ``scaling pattern"~\cite{bj69} in the nucleon structure
functions, consistent with scattering from partons carrying the quantum
numbers of the Gell-Mann/Zweig 
quarks.  Further experimental studies of muon-nucleon and neutrino-nucleon
deep inelastic scattering experiments at CERN and Fermilab established
beyond any doubt the quark-parton model~(QPM) of the
nucleon~\cite{cl79}, and provided substantial supporting evidence for the
emerging theory of quantum chromodynamics~(QCD).

The cross section for inelastic electron-nucleon scattering is given in
terms of the structure functions $F_1(\nu,Q^2)$ and $F_2(\nu,Q^2)$ of the
nucleon by:
\begin{equation}
\label{eqsig}
{  {\sigma} \equiv 
 { {d^2\sigma} \over {d\Omega dE'} }(E,E',\theta)  =
 { {4 \alpha^2 (E')^2} \over {Q^4} }  \cos^2 \left( {\theta \over 2} \right)
\left[ { F_2(\nu,Q^2) \over \nu } + 
 { {2 F_1(\nu,Q^2)} \over {M} } \tan^2 \left( {\theta \over 2} \right) \right]\ ,  }
\end{equation}
where $\alpha$ is the fine-structure constant, $E$ is the incident electron 
energy, $E'$ and $\theta$ are the scattered electron energy and angle, 
$\nu = E-E'$ is the energy transfer, $Q^2=4 E E' \sin^2(\theta/2)$ is
minus the four-momentum transfer squared, and $M$ is the nucleon mass. 

The basic idea of the quark-parton model~\cite{bj69,fe69} is to represent
inelastic electron-nucleon scattering as quasi-free scattering from 
the partons/quarks in the nucleon, when viewed in a frame where the nucleon
has infinite momentum (the center-of-mass frame is a very good
approximation to such a frame).   The fractional momentum of the nucleon
carried by the struck quark is given by the Bjorken scaling variable,
$x=Q^2/2M\nu$. In the limit where $\nu \rightarrow \infty$,
$Q^2 \rightarrow \infty$ with $x$ fixed, the nucleon structure functions
become:
\begin{equation}
\label{eqf12} 
F_1 = {1 \over 2} \sum_i e_i^2 f_i(x)~ , ~~~~~~~~ F_2 = x \sum_i
e_i^2f_i(x)\ .
\end{equation}
Here, $e_i$ is the fractional charge of quark type $i$, $f_i(x)dx$ is 
the probability that a quark of type $i$ carries momentum in the range
between $x$ and $x+dx$, and the sum runs over all quark types.

Since the charges of the $u,d$ and $s$ quarks are 2/3, $-1/3$ and
$-1/3$, respectively, the $F_2(x)$ structure function for the proton is
given by:                 
\begin{equation}
\label{eqf2p}
F^p_2(x) = x \left[ \left({2 \over 3} \right)^2 \left(u+\bar{u} \right)+
\left( -{1 \over 3} \right)^2 \left(d+\bar{d} \right) +
\left( -{1 \over 3} \right)^2 \left(s+\bar{s}\right) \right].
\end{equation} 
The structure function of the neutron
is related to that of the proton by isospin symmetry. 
Since the up/down quarks and proton/neutron both form isospin doublets
one has:
$u^p(x) = d^n(x) \equiv u(x)$, $d^p(x) = u^n(x) \equiv d(x)$,
$s^p(x) = s^n(x) \equiv s(x)$ (with analogous relations for the
antiquarks), and:  
\begin{equation}
\label{eqf2n}
F^n_2(x) = x \left[ \left(-{1 \over 3} \right)^2 \left( u+\bar{u} \right)+
\left( {2 \over 3} \right )^2 \left( d+\bar{d} \right) +
\left( -{1 \over 3} \right )^2 \left( s+\bar{s} \right) \right]\ .
\end{equation} 
%
Equations 3 and 4 result in the structure function ratio:
\begin{equation}
\label{eqfnp}
{F^n_2 \over F^p_2} = { {  (u + \bar{u}) + (s + \bar{s})  + 4 (d + \bar{d}) } 
\over { 4 (u + \bar{u}) +  (d + \bar{d}) + (s + \bar{s})  } }\ .
\end{equation}
Since all the distribution functions must be positive for all $x$, the         
above expression is bound for all $x$ by:
\begin{equation}
\label{eqnac}
{1 \over 4} \leq {F^n_2 \over F^p_2 } \leq 4\ ,
\end{equation}
which is known as the Nachtmann inequality~\cite{na72}.
Figure 1 (left) shows all the SLAC deep inelastic scattering (DIS)
(large $Q^2$ and $\nu$) data from the pioneering SLAC/MIT
Collaboration experiments on the $F_2^n/F_2^p$ ratio versus $x$.         
The ratio has been extracted from DIS measurements off the proton
and deuteron~\cite{bo73}, using a non-relativistic smearing model to account 
for the Fermi-motion
of the nucleons in the deuteron~\cite{bo79}.  The ratio data are within
the bounds of the Nachtmann inequality.
For large $x$, the ratio is about 1/4 which can only be reached if       
$d = \bar{d} = s = \bar{s} = 0$. This suggests a picture in which the 
high momentum partons in the proton (neutron) are mainly up (down) quarks.
For small $x$ the ratio is close to 1, suggesting little influence             
of valence quarks and dominance of the quark-antiquark ``sea''.

\begin{figure}[t]
\begin{center}
\epsfig{file=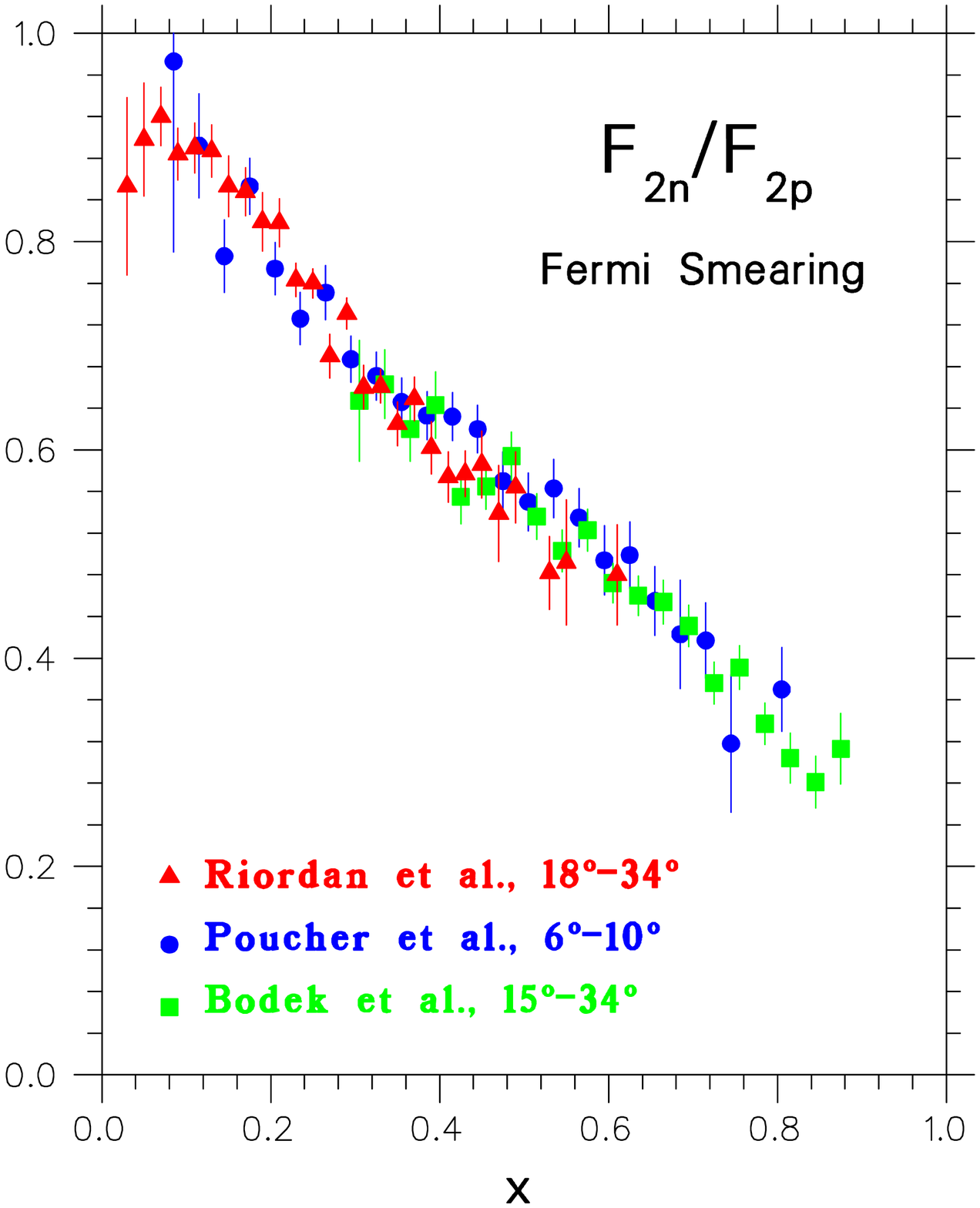,width=3.2in}
\psfig{file=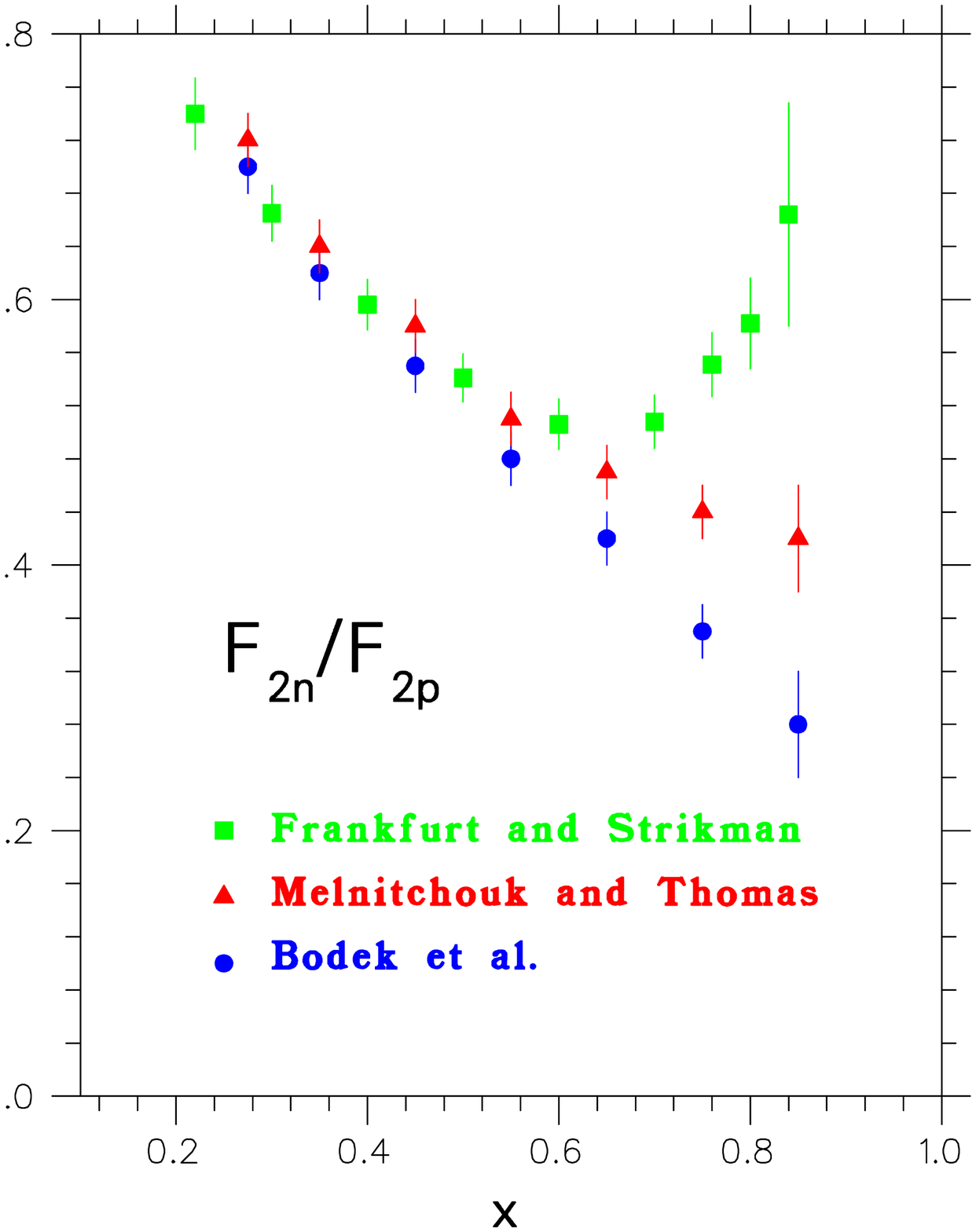,width=3.13in}
\caption{\small 
Left: SLAC data~\cite{bo73} on the $F_2^n/F_2^p$ ratio extracted
         from proton and deuteron DIS measurements with a 
         Fermi-smearing model~\cite{bo79}.         
Right: The $F_2^n/F_2^p$ ratio extracted
         from proton and deuteron DIS measurements~\cite{bo73} with 
         a) a Fermi-smearing model (Bodek {\it et al.}~\cite{bo79}),
         b) a covariant model that includes binding and off-shell effects 
            (Melnitchouk and Thomas~\cite{me96}), 
            and c) the ``nuclear density model'' that also incorporates
            binding effects (Frankfurt and Strikman~\cite{ya99,wh92,fr88}). 
         }
\end{center}
\vskip-0.5cm
\end{figure}

\section{Theory Overview}

The $F_2^n/F_2^p$ ratio can be calculated in a number of models of the
nucleon.
In a world of exact SU(6) symmetry, the wave function of a proton,
polarized in the $+z$ direction for instance, would be simply~\cite{cl79}:
\begin{eqnarray} 
\label{eqsu6}
p\uparrow 
&=& {1 \over \sqrt{2}}  u\uparrow (ud)_{S=0}\ 
 +\ {1 \over \sqrt{18}} u\uparrow (ud)_{S=1}\ 
 -\ {1 \over 3}         u\downarrow (ud)_{S=1}\ \nonumber \\ 
& & 
 -\ {1 \over 3}         d\uparrow (uu)_{S=1}\ 
 -\ {\sqrt{2} \over 3}  d\downarrow (uu)_{S=1}\ , 
\end{eqnarray} 
where the subscript $S$ denotes the total spin of the ``diquark''
partner of the quark.
In deep-inelastic scattering, exact SU(6) symmetry would be 
manifested in equivalent shapes for the valence quark  
distributions of the proton, which would be related simply by 
$u_v(x) = 2 d_v(x)$ for all $x$. 
For the neutron to proton structure function ratio this would
imply~\cite{ku71}:
\begin{eqnarray}
{ F_2^n \over F_2^p } 
&=& {2 \over 3}, \ \ \ \ \ \ 
{d \over u} = {1 \over 2}
\ \ \ \ \ {\rm [SU(6)\ symmetry]}.
\end{eqnarray}

In nature, spin-flavor SU(6) symmetry is, of course, broken. 
The nucleon and $\Delta$ masses are split by some 300 MeV.    
In deep inelastic scattering off the nucleon, this symmetry breaking is  
reflected in the experimental observation that the $d$ quark
distribution is softer than the $u$ quark distribution, with the 
$F_2^n/F_2^p$ ratio deviating from the SU(6) expectation.  
The correlation between the mass splitting in the {\bf 56} baryons  
and the large-$x$ behavior of $F_2^n/F_2^p$ was observed some  
time ago by Close~\cite{cl73} and Carlitz~\cite{ca75}.    
Based on phenomenological~\cite{cl73} and Regge~\cite{ca75} considerations,
and kinematic arguments by Close and Thomas~\cite{cl88},
the breaking of the symmetry in Equation \ref{eqsu6} was
argued to arise from a suppression of the ``diquark'' configurations 
having $S=1$ relative to the $S=0$ configuration as $x \rightarrow 1$.  
From Equation \ref{eqsu6}, a dominant scalar valence diquark component 
of the proton suggests that in the $x \rightarrow 1$ limit, $F_2^p$  
is essentially given by a single quark distribution (i.e. the $u$),  
in which case:  
\begin{eqnarray} 
{ F_2^n \over F_2^p } 
&\rightarrow& { 1 \over 4 }, \ \ \ \ \  
{ d \over u } \rightarrow 0\ \ \ \ \ 
[S=0\ {\rm dominance}]. 
\end{eqnarray} 
This expectation has, in fact, been built into most phenomenological  
fits to the parton distribution data~\cite{ei84}. 
 
The phenomenological suppression of the $d$ quark distribution
at intermediate and large $x$ 
can be understood in terms of the same hyperfine interaction 
responsible for the $N-\Delta$ splitting~\cite{cl88,is78,is99}. 
This color hyperfine interaction is generated by one-gluon exchange
between quarks in the core~\cite{is78}.
At lowest order, the Hamiltonian for the color-magnetic hyperfine
interaction~\cite{is78} between two quarks is proportional to
$\vec S_i \cdot \vec S_j$, where $\vec S_i$ is the spin vector
of quark $i$.
Because this force is repulsive if the spins of the quarks are parallel
and attractive if they are antiparallel, Isgur~\cite{is99} argued
that the SU(6) wave function
in Equation \ref{eqsu6} naturally leads to an increase in the mass of
the $\Delta$ and a lowering of the mass of the nucleon, and a softening
of the $d$ quark distribution relative to the $u$.

An alternative suggestion, based on a perturbative QCD argument, 
was originally formulated by Farrar and Jackson~\cite{fa75}.
Their proposal might be expected to hold at very large $x$, where
the struck quark has momentum approaching infinity, which goes
beyond mean field theory (like the simple quark model).
They showed that the exchange of longitudinal gluons,
which are the only type permitted when the spins of the two quarks   
in $(qq)_S$ are aligned, would introduce a factor $(1-x)^{1/2}$ into  
the Compton amplitude -- in comparison with the exchange of a  
transverse gluon between quarks with spins anti-aligned. 
In this approach the relevant component of the proton valence  
wave function at large $x$ is that associated with states in  
which the total ``diquark'' spin projection, $S_z$, is zero
as $x \rightarrow 1$.
Consequently, scattering from a quark polarized in the opposite 
direction to the proton polarization is suppressed by a factor  
$(1-x)$ relative to the helicity-aligned configuration. 

A similar result is also obtained in the treatment of  
Brodsky {\em et al.}~\cite{bo95} (based on quark-counting rules), 
where the large-$x$ behavior of the  
parton distribution for a quark polarized parallel ($\Delta S_z = 1$) 
or antiparallel ($\Delta S_z = 0$) to the proton helicity is given by:  
$q^{\uparrow\downarrow}(x) =  (1-x)^{2n-1+\Delta S_z}$,
where $n$ is the minimum number of non-interacting quarks (equal  
to 2 for the valence quark distributions). 
Using Equation \ref{eqsu6}, in the $x \rightarrow 1$ limit one therefore
predicts: 
\begin{eqnarray} 
{ F_2^n \over F_2^p } 
&\rightarrow& {3 \over 7}, \ \ \ \ \  
{ d \over u } \rightarrow { 1 \over 5 }\ \ \ \ \ 
[S_z=0\ {\rm dominance}]. 
\end{eqnarray} 
It should be noted that in the latter two treatments, the $d/u$ ratio 
does not vanish and the $F_2^n/F_2^p$ ratio is 3/7 instead of 1/4. 

\vskip0.4in
\section{Motivation for a New Experiment}

Although the problem of extracting neutron structure functions from
deuteron data is rather old~\cite{we71}, the discussion has been
recently revived~\cite{me96,ya99} with the realization~\cite{wh92}
that $F_2^n$, extracted from
$F_2^d$ by taking into account Fermi-motion {\em and} binding
effects, could be significantly larger~\cite{me96,ya99,wh92} than that
extracted in earlier
analyses~\cite{bo79} in which only Fermi-motion corrections were applied.

Melnitchouk and Thomas~\cite{me96} have incorporated binding and off-shell
effects within a covariant framework in terms of relativistic deuteron
wave functions (as calculated by Gross and collaborators~\cite{gr79},
for instance).
Neglecting the relativistic deuteron $P$-states and off-shell deformation
of the bound nucleon structure function (which were found to contribute at
the $\sim 1\%$ level~\cite{me96}), the deuteron structure function,
$F_2^d$, can be written as a convolution of the free proton and neutron
$F_2$ structure functions and a nucleon momentum distribution in the
deuteron, $f_{N/d}$:
\begin{equation}
\label{eqdcon}
F_2^d(x,Q^2)
  =  \int { dy\  f_{N/d}(y)\ [ F_2^p(x/y,Q^2) + F_2^n(x/y,Q^2) ] },
\end{equation}
where $y$ is the fraction of the `plus'-component of the nuclear momentum
carried by the interacting nucleon, and $f_{N/d}(y)$ takes into account
both Fermi-motion and binding effects.
Their calculation results in larger $F_2^n/F_2^p$ values compared with the
Fermi-motion only extracted values.
As can be seen in Figure 1 (right), the difference at $x=0.85$ can be up to
$\sim 50\%$.

Whitlow {\it et al.}~\cite{wh92} incorporated binding effects by
estimating the EMC effect~\cite{au83,go94} of the deuteron itself
(difference between the free and bound nucleon quark distributions)
using the ``nuclear density model'' of Frankfurt and Strikman~\cite{fr88}.
In this model, the EMC effect for the deuteron scales with nuclear density
as for heavy nuclei:
\begin{equation}
\label{junk}
 \frac{F_2^d}{F_2^p +F_2^n} = 1 + 
\frac{\rho_d}{\rho_A - \rho_d}\left [ \frac{F_2^A}{F_2^d} -1 \right ]\ ,
\end{equation}
where $\rho_d$ is the deuteron charge density, and $\rho_A$ and $F_2^A$
refer to a heavy nucleus with mass number $A$.  This model predicts
$F_2^n/F_2^p$ values that are significantly higher ($> 100\%$) than the
Fermi-motion only extracted ones at high $x$, as can be seen in Fig. 1
(right).

It is evident from the above two models that neglecting
nuclear binding effects in the deuteron can introduce, at large $x$, a significant 
uncertainty in the extraction of $F_2^n/F_2^p$ and $d/u$.
In the absence of experimental data or a unique theory for the magnitude
of binding effects and the existence of the EMC effect in the deuteron,
the question of the large-$x$ behavior of $F_2^n/F_2^p$ and $d/u$ can
only be settled by a measurement which does not rely on the use of the
deuteron as an effective neutron target.

The above situation can be remedied by using a method proposed by
Afnan {\it et al.}~\cite{af00}, which maximally exploits the mirror
symmetry of $A=3$ nuclei and extracts the $F_2^n/F_2^p$ ratio from DIS
measurements off $^3$H and $^3$He.
Regardless of the absolute values of the nuclear EMC effects in $^3$He or 
$^3$H, the differences between these will be small -- on the scale of 
charge symmetry breaking in the nucleus -- which allows a determination of
the $F_2^n/F_2^p$ and $d/u$ ratios at large-$x$ values essentially free of nuclear
contamination.

\section{Exploring Deep Inelastic Scattering off $^3$H and $^3$He}

In the absence of a Coulomb interaction and in an isospin symmetric world, 
the properties of a proton (neutron) bound in the $^3$He nucleus would be 
identical to that of a neutron (proton) bound in the $^3$H nucleus. 
If, in addition, the proton and neutron distributions in $^3$He 
(and in $^3$H) were identical, the neutron structure function could 
be extracted with no nuclear corrections, regardless of the 
size of the EMC effect in $^3$He or $^3$H separately. 
In practice, $^3$He and $^3$H are of course not perfect mirror nuclei 
-- their binding energies for instance differ by some 10\% -- and the 
proton and neutron distributions are not quite identical. 
However, the $A=3$ system has been studied for many years, and modern 
realistic $A=3$ wave functions are known to rather good accuracy. 

Defining the EMC-type ratios for the $F_2$ structure functions of 
$^3$He and $^3$H (weighted by corresponding isospin factors) by: 
%
\begin{eqnarray} 
R(^3{\rm He}) = { F_2^{^3{\rm He}} \over 2 F_2^p + F_2^n }~,~~~~~~~~~ 
%
R(^3{\rm H}) = { F_2^{^3{\rm H}} \over F_2^p + 2 F_2^n }\ , 
\end{eqnarray}   
%
one can write the ratio of these as 
${\cal R}^* = R(^3{\rm He})/R(^3{\rm H})$, which directly yields 
%
%
the ratio of the free
neutron to proton structure functions: 
\begin{eqnarray} 
\label{Eqnp} 
{ F_2^n \over F_2^p } 
&=& { 2 {\cal R}^* - F_2^{^3{\rm He}}/F_2^{^3{\rm H}} 
\over 2 F_2^{^3{\rm He}}/F_2^{^3{\rm H}} - {\cal R}^* }\ . 
\end{eqnarray} 
The $F_2^n/F_2^p$ ratio extracted via Equation \ref{Eqnp} does not depend
on the size of the EMC effect in $^3$He or $^3$H, but rather on the 
{\em ratio} of the EMC effects in $^3$He and $^3$H. 
If the neutron and proton distributions in the $A=3$ nuclei are not 
dramatically different, one might expect ${\cal R}^* \approx 1$. 
To test whether this is indeed the case requires an explicit calculation 
of the EMC effect in the $A=3$ system. 

The conventional approach employed in calculating nuclear structure  
functions in the valence quark region is the impulse 
approximation, in which the virtual photon, $\gamma^*$, scatters incoherently from 
individual nucleons in the nucleus~\cite{ge95}. 
The nuclear cross section is determined by factorizing the 
$\gamma^*$--nucleus interaction into $\gamma^*$--nucleon and 
nucleon--nucleus amplitudes. 
The structure function of a nucleus, $F_2^A$, 
can then be calculated by folding the nucleon structure function, 
$F_2^N$, with the nucleon momentum distribution in the nucleus, $f_{N/A}$: 
\begin{eqnarray} 
\label{con} 
F_2^A(x) &=& \int dy\ f_{N/A}(y)\ F_2^N(x/y)\ 
\equiv\ f_{N/A}(x) \otimes F_2^N(x)\ , 
\end{eqnarray} 
where the $Q^2$ dependence in the structure functions is implicit. 
The convolution expression in Equation \ref{con} is correct in the limit of 
large $Q^2$; at finite $Q^2$ there are additional contributions to 
$F_2^A$ from the nucleon $F_1^N$ structure functions, although these are 
suppressed by powers of $M^2/Q^2$, where $M$ is the nucleon mass. 

The distribution $f(y)$ of nucleons in the nucleus is related to the 
nucleon spectral function $S(p)$ by \cite{ge95}: 
\begin{eqnarray} 
f(y) = \int d^3\vec p\ 
\left( 1 + {p_z \over p_0} \right) 
\delta \left( y - { p_0 + p_z \over M } \right) S(p)\ , 
\end{eqnarray} 
where $p$ is the momentum of the bound nucleon.
For an $A=3$ nucleus, $S(p)$ is evaluated from the 
three-body nuclear wave function, calculated by either solving the
homogeneous Faddeev equation with a given two-body interaction
\cite{af00,bi00} or by using a variational technique \cite{ciofi}.
Details of the computation of the wave functions can be found in 
Ref. \cite{bi00}.

In terms of the proton and neutron momentum distributions, the $F_2$ 
structure function for $^3$He is given by: 
\begin{equation}
F_2^{^3{\rm He}}\ 
= 2\ f_{p/^3{\rm He}}\ \otimes\ F_2^p\ 
 +\    f_{n/^3{\rm He}}\ \otimes\ F_2^n\ .
\end{equation} 
%
%
Similarly for $^3$H, the structure function is evaluated from the proton 
and neutron momentum distributions in $^3$H:  
\begin{equation}
F_2^{^3{\rm H}} 
 =    f_{p/^3{\rm H}} \otimes\ F_2^p\ 
 +\ 2\ f_{n/^3{\rm H}} \otimes\ F_2^n\ .
\end{equation}  
%
%
Because isospin symmetry breaking effects in nuclei are quite small, 
one can to a good approximation relate the proton and neutron 
distributions in $^3$He to those in $^3$H: 
\begin{equation}
f_{n/^3{\rm H}} \approx f_{p/^3{\rm He}}~, ~~~~~~~~
f_{p/^3{\rm H}} \approx f_{n/^3{\rm He}}\ ,
\end{equation} 
%
although in practice one has to consider both the isospin symmetric and isospin 
symmetry breaking cases explicitly. 

The ratio ${\cal R}^*$ of EMC ratios for $^3$He and $^3$H, as calculated
by Afnan {\it et al.}~\cite{af00} is shown in 
Figure 2 (left) for nuclear model wave functions based on a) the
``Paris~(EST)'' separable approximation~\cite{ha84} to the Paris potential,
b) the unitary pole approximation~\cite{af73} of the Reid Soft Core (RSC)
potential, and c) the Yamaguchi potential with 7\%
mixing between $^3 S_1$ and $^3 D_1$ waves~\cite{ya54}.
In all three cases, the CTEQ parameterization~\cite{ei84} of parton 
distributions at $Q^2=10$~(GeV/$c$)$^2$ was used for $F_2^N$. 
The EMC effects are seen to largely cancel over a large range of $x$, 
out to $x \sim 0.9$, with a deviation from unity of less than 2\%. 
Furthermore, the dependence on the nuclear wave function is very weak.
The pattern of behavior of the ratio ${\cal R}^*$ has been confirmed
in independent calculations by Liuti using the formalism of
Ref.~\cite{liuti}, and by Pace {\it et al.}~\cite{pace}, using a
variational approach to calculate the three-body spectral function. 

\begin{figure}[t]
\begin{center}
\epsfig{file=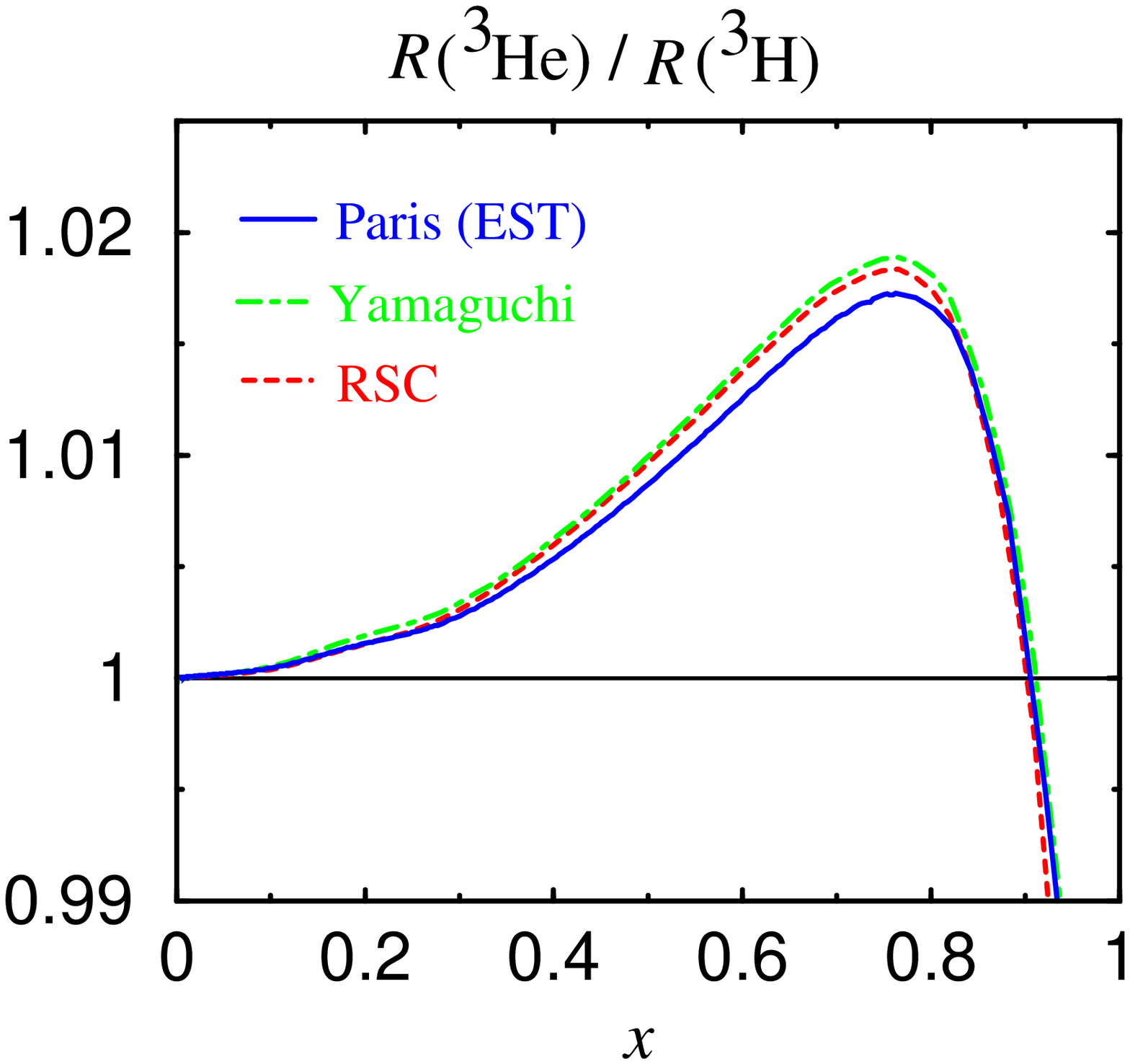,width=3.2in}
\epsfig{file=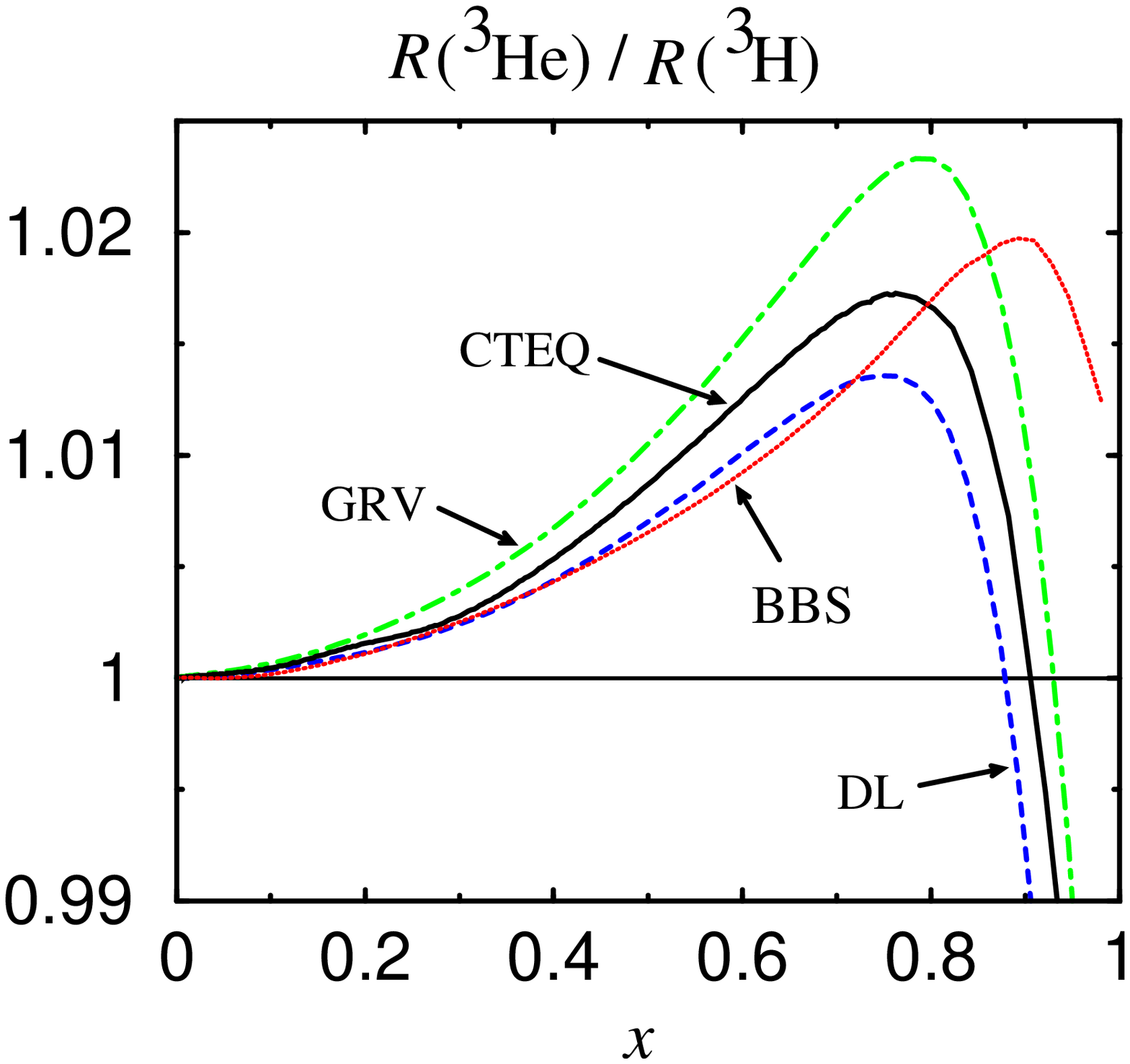,width=3.2in}
\caption{\small 
Left:   Ratio of nuclear EMC ratios for $^3$He and $^3$H for
        various nuclear model wavefunctions as calculated by Afnan 
        {\it et al.}~\cite{af00} using the CTEQ nucleon structure
        function parametrization (see text).
Right:  Ratio of nuclear EMC ratios for $^3$He and $^3$H with the
        Paris~(EST) wave functions, using various nucleon structure function
        parameterizations~\cite{af00} (see text).        
        }
\end{center}
\vskip-0.5cm
\end{figure}

The dependence of ${\cal R}^*$ on the input nucleon structure function 
parameterization is illustrated in Figure 2 (right), where several representative 
curves at $Q^2 = 10$ (GeV/$c$)$^2$ are given: apart from the standard CTEQ 
fit (solid), the results for the GRV~\cite{gl98} (dot-dashed), 
Donnachie-Landshoff (DL)~\cite{do94} (dashed), 
and BBS~\cite{bo95} (dotted) parameterizations are also shown 
(the latter at $Q^2=4$~(GeV/$c$)$^2$). 
Despite the seemingly strong dependence on the nucleon structure function 
input at very large $x$, this dependence is actually artificial.  
In practice, once the ratio $F_2^{^3{\rm He}}/F_2^{^3{\rm H}}$ is 
measured, one can employ an iterative procedure to eliminate this 
dependence altogether. 
Namely, after extracting $F_2^n/F_2^p$ from the data using some 
calculated ${\cal R}^*$, the extracted $F_2^n$ can then be used to 
compute a new ${\cal R}^*$, which is then used to extract a new and 
better value of $F_2^n/F_2^p$. 
This procedure is iterated until convergence is achieved and a
self-consistent solution for the extracted $F_2^n/F_2^p$ is obtained.
Both Afnan {\it et al.}~\cite{af00} and Pace {\it et al.}~\cite{pace} have
independently confirmed the convergence of this procedure.
 
All of the above suggest that, for the purpose of this investigation, 
it is reasonable and safe to assume that we can describe ${\cal R}^*$ with a central
value and assign a systematic uncertainty that grows from 0\% at $x=0$
to $\pm$1\% at $x=0.82$.  Further theoretical investigations in the future
could possibly reduce this uncertainty. 

\section{The Experiment}  
 
The proposed energy upgrade of Jefferson Lab (JLab) will offer a unique opportunity 
to perform deep inelastic scattering off the $^3$He and $^3$H mirror 
nuclei at large-$x$ values.   
The DIS cross section for $^3$H and $^3$He is given in terms of their 
$F_1$ and $F_2$ structure functions by Equation 1, where $M$  
represents in this case the nuclear mass.  The nuclear  
structure functions $F_1$ and $F_2$ are connected through the ratio 
$R = {\sigma_L / \sigma_T}$, where $\sigma_L$ and $\sigma_T$ are the  
virtual photoabsorption 
cross sections for longitudinally and transversely polarized photons,
by $F_1 = { [{F_2 (1 + Q^2 /\nu^2)}] /[{2 x (1+R)}] }$. 
The ratio $R$ has been measured to be independent of the mass
number, $A$, in precise SLAC and CERN measurements using hydrogen,  
deuterium, iron and  other nuclei (for a compilation of data see 
Ref. \cite{ge95}). 

By performing the tritium and helium measurements under identical 
conditions, using the same incident beam and scattered electron  
detection system configurations (same $E$, $E'$ and $\theta$),  
and assuming that the ratio $R$ is the same for both nuclei, the ratio 
of the DIS cross sections for the two nuclei will provide a direct  
measurement of the ratio of their $F_2$ structure functions: 
$ {\sigma(\rm ^3H)} / {\sigma(\rm ^3He)} = 
{ F_2  (\rm ^3H)} / {F_2   (\rm ^3He)} $.

The key issue for this experiment will be the availability of a 
high density tritium target.  Tritium targets have been used 
in the past to measure the elastic form factors of $\rm ^3$H at 
Saclay~\cite{am94} and MIT-Bates~\cite{be89}.  The Saclay target  
contained liquid $^3$H at 22~K  and was  
able to tolerate beam currents up to 10~$\mu$A 
with very well understood beam-induced density changes.  The 
tritium density (0.271 g/cm$^3$) at the operating conditions  
of this target was known to $\pm 0.5 \%$ (based on actual 
density measurements).   
The MIT-Bates target contained gas $^3$H at 45~K/15~atm and 
was able to tolerate beam currents up to 25~$\mu$A with 
measurable small density changes.  The tritium 
density, under these operating conditions, has been determined 
to be 0.028 g/cm$^3$ with $\sim\pm 2\%$ uncertainty, using the
Virial formalism.  

Given a tritium target, an entire program of elastic, quasielastic and
inelastic measurements will be possible at JLab.  This program can be
better accomplished by building a target similar to the MIT-Bates one
(the cooling mechanism of a target similar to the Saclay one would
prevent coincidence measurements).  The tritium density
can be better determined from comparison of the elastic cross section
measured with the 45~K/15~atm cell and a cell filled up with
tritium at higher temperatures (ideal gas of known density).
Two more cells will also be necessary for the $^3$He measurements.  

The availability of the proposed Medium Acceptance Device (MAD) JLab Hall A
spectrometer~\cite{mad} will facilitate high statistics DIS cross section 
measurements ($\le \pm 0.25 \%$) in a large-$x$ range as well as valuable
systematics checks.  The performance of the
above spectrometer is expected to be comparable, if not better, to that
of the SLAC 8 GeV/c spectrometer that has provided precise measurements
for absolute DIS cross sections, DIS cross section ratios, and differences
in $R$ for several nuclei~\cite{go94,da94,tao}.  The overall systematic
errors for these measurements have been typically $\pm2\%$, $\pm0.5\%$ and
$\pm 0.01$, respectively.
Since the objective of the experiment is based on measurements of 
cross section ratios, many of 
the experimental errors that plague absolute cross section measurements will 
cancel out.  The experimental uncertainties on the ratio of cross 
sections should be similar to those achieved by SLAC experiments 
E139~\cite{go94} and E140~\cite{da94,tao}, which were typically around
$\pm0.5\%$. 

\begin{figure}[t]
\begin{center}
\psfig{file=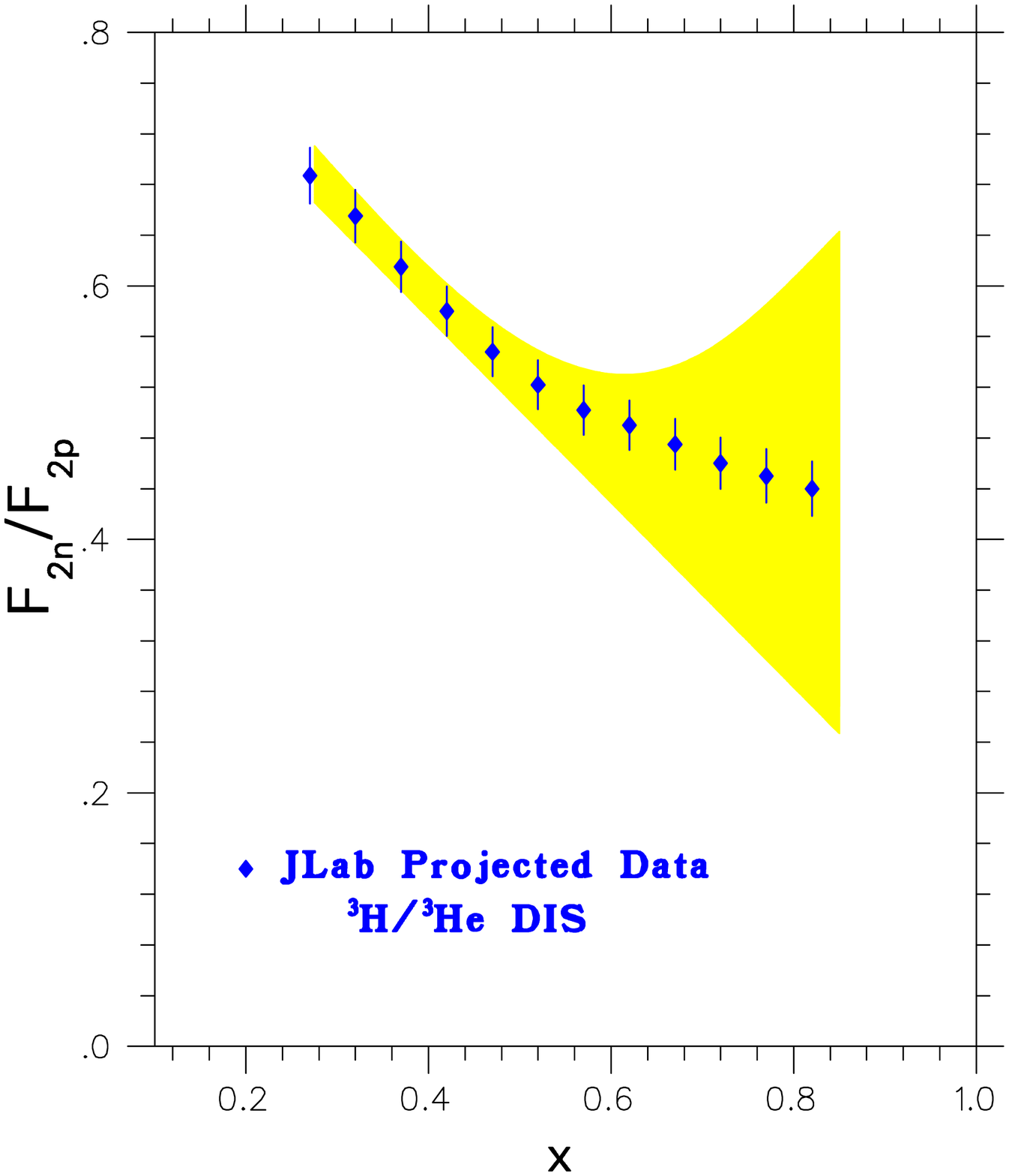,width=3.2in}
\psfig{file=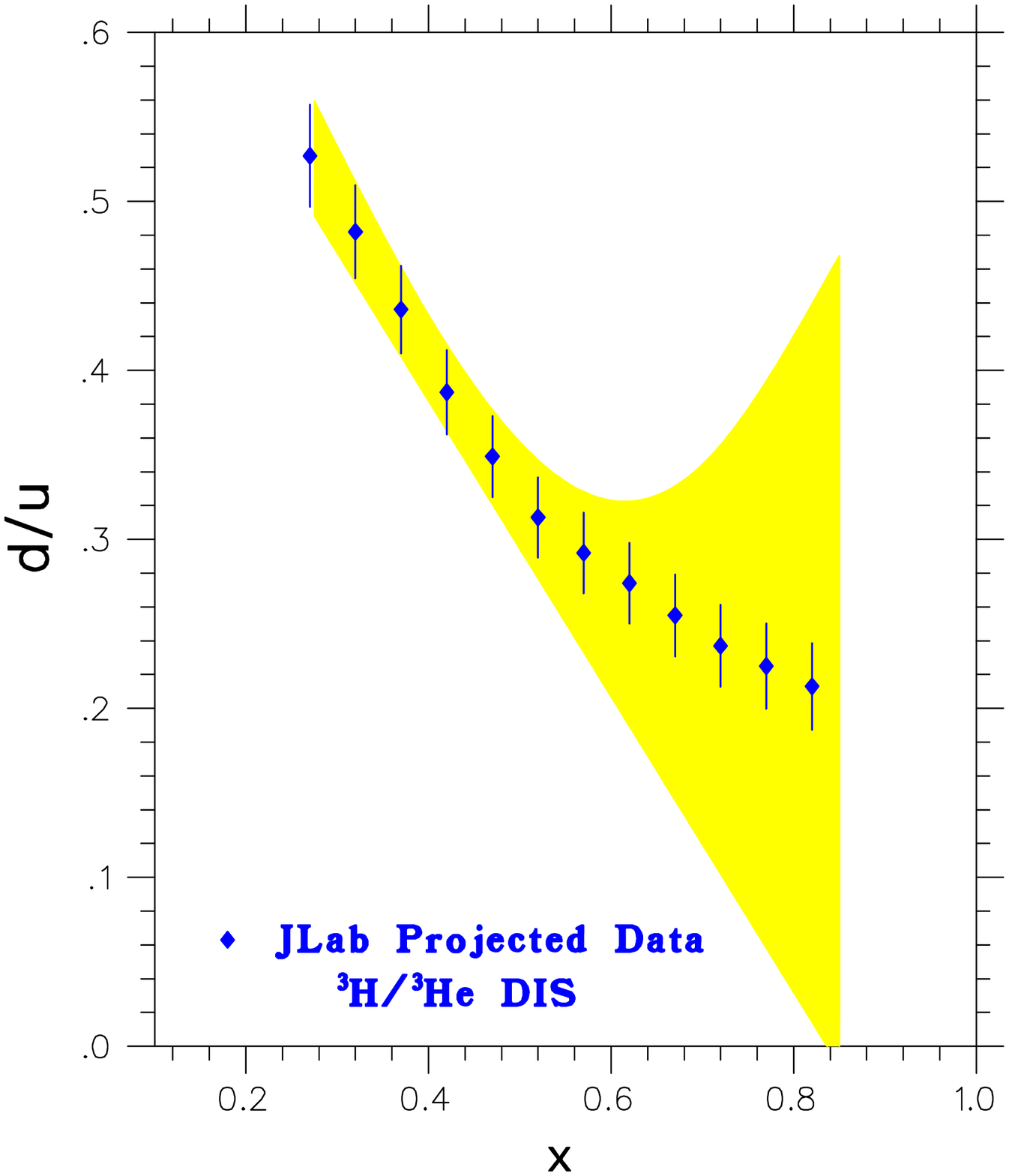,width=3.2in}
\caption{\small Projected data for the $F_2^n/F_2^p$ structure function
         (left) and $d/u$ quark (right) ratios
         from the proposed $^3$H/$^3$He JLab DIS experiment.
         The error bars include experimental and theoretical
         systematic uncertainties added in quadrature.  The
         shaded band indicates the present uncertainty due
         to possible binding effects in deuteron.}
\end{center}
\vskip-0.5cm
\end{figure}

DIS scattering with the proposed 11 GeV JLab electron beam can provide 
measurements 
for the $^3$H and $^3$He $F_2$ structure functions in the $x$ range 
from 0.10 to 0.82.  The electron scattering angle will range from 
12$^\circ$ to 47$^\circ$ and the electron scattered energy from 
1 to 4 GeV.  
Assuming $^3$H and $^3$He luminosities of 
$\sim 5 \times 10^{37}$ cm$^{-2}$ s$^{-1}$, the time required for 
the above ``core'' set of measurements has been estimated to be
less than a week.  The DIS cross section has been estimated to
be between 0.01 and 100 nb/sr/GeV assuming 
that $\sigma(^3He) \simeq 2 \sigma_p + \sigma_n$ and  
$\sigma(^3H) \simeq 2 \sigma_d - \sigma_p$, and using values for the  
proton ($\sigma_p$) and deuteron ($\sigma_d$) DIS cross sections  
and the ratio $R$ from the SLAC ``global'' analysis~\cite{wh92}. 
It is evident that such an experiment will be able to provide very  
high statistics data and perform necessary systematic studies in  
a very timely fashion. 

\begin{figure}[t]
\begin{center}
\psfig{file=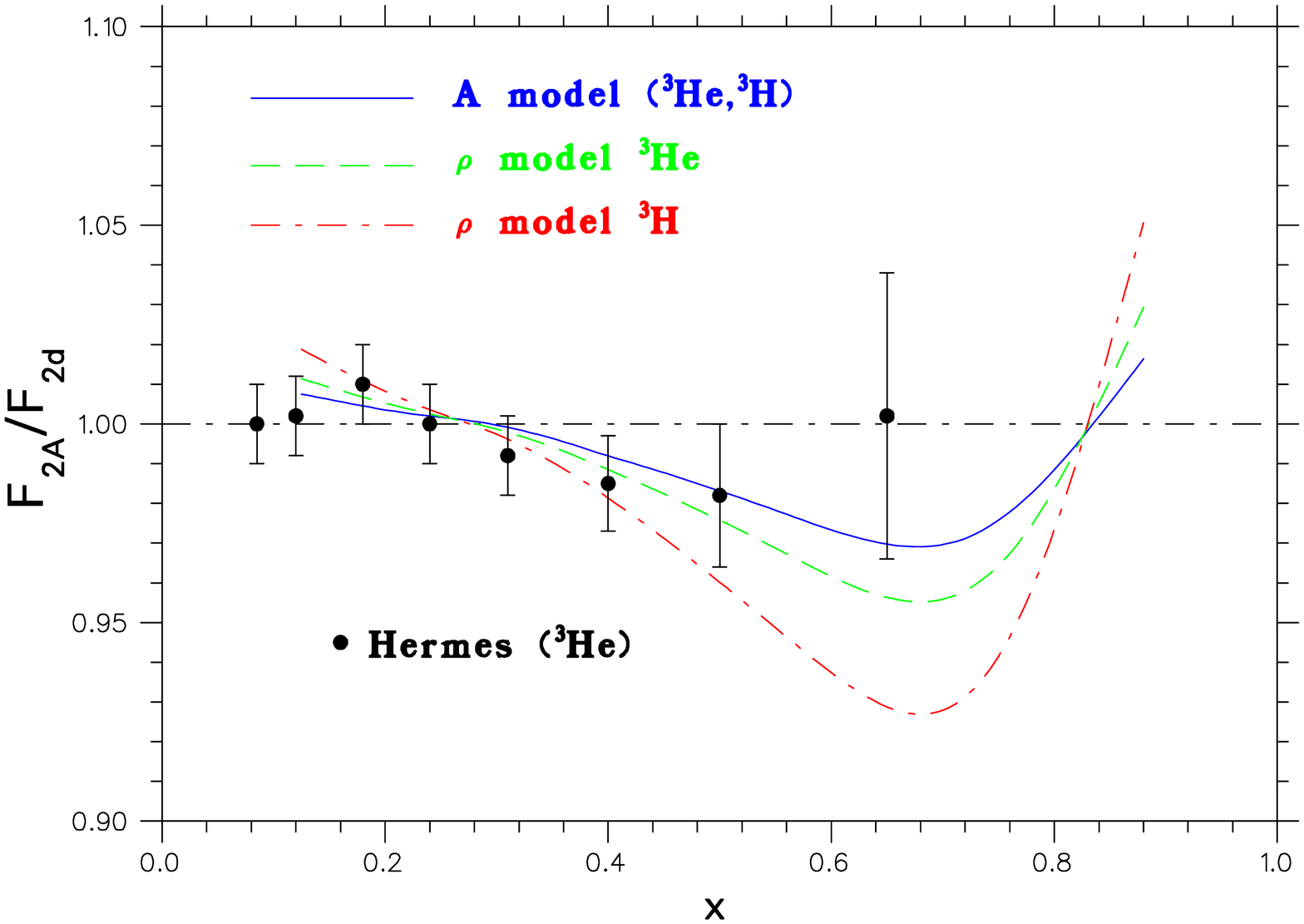,height=4.4in,angle=90}
\caption{\small The $^3$H and $^3$He isoscalar EMC effect ratios
         $F_2(^3H)/F_2(d)$ and $F_2(^3He)/F_2(d)$
         as predicted~\cite{go94} by the atomic mass $A$ model  
         and the nuclear density $\rho$ model. Also shown are recent
         data from the Hermes/DESY experiment~\cite{ac00}.}
\end{center}
\vskip-0.8cm
\end{figure}

An important systematic check will be to confirm that the ratio $R$
is the same for $^3$H and $^3$He.  The 11 GeV beam and the momentum
and angular range of MAD will allow measurements of $R$ in the 
same $x$ range as in the SLAC experiments by means of a
Rosenbluth separation.
The $R$ measurements will be limited by 
inherent systematics uncertainties rather than statistics as
in the SLAC case.  
It is estimated that the $R$ measurements will require an amount
of beam time comparable to the one required for the above core set of
measurements.

The $F_2(^3\rm H)/F_2(^3\rm He)$ ratio is expected to be 
dominated by experimental uncertainties that do not cancel in 
the DIS cross section ratio of $^3\rm H$ to $^3\rm He$ and  
the theoretical uncertainty in the calculation of the ratio ${\cal R}^*$. 
Assuming that the target densities can be known to the $\pm0.5\%$ 
level and that the 
relative difference in the $^3$H and $^3$He radiative corrections 
would be $\pm0.5\%$ as in Refs~\cite{go94,da94}, the total experimental 
error in the the DIS cross section ratio of $^3\rm H$ to $^3\rm He$ 
should be $\sim\pm1.0\%$.  Such an error is comparable to a realistic
maximum theoretical uncertainty ($\sim\pm1\%$ in the vicinity of $x$~=~0.8)
in the calculation of the ratio ${\cal R}^*$. 

The quality of the projected data for the $F_2^n/F_2^p$ and $d/u$
ratios is shown in Figure 3.  The error bars assume
a $\pm1\%$ overall systematic experimental error in the measurement 
of the $\sigma(^3H)/\sigma(^3He)$ ratio and a theoretical
uncertainty in ${\cal R}^*$ that increases linearly from 0\% at
$x=0$ to $\pm1\%$ at $x=0.82$.  The shaded areas in Fig. 3
indicate the present uncertainty due to possible nuclear corrections 
in the extraction of $F_2^n/F_2^p$ and $d/u$ from deuterium data. 
It is evident that the proposed measurement will be able to 
unquestionably distinguish between the present competing 
extractions of the $F_2^n/F_2^p$ and $d/u$ ratios from proton and 
deuterium DIS measurements, 
and determine their values with an 
unprecedented precision in an almost model-independent way.  

A secondary goal of this experiment is 
the precise determination of the EMC effect in $^3$H and $^3$He. 
At present time, the precision of the available SLAC and CERN data allow 
for two equally compatible parametrizations~\cite{go94} of the EMC effect.  
In the first one, the EMC effect is parametrized versus the 
mass number $A$ and in the second one versus the nuclear density 
$\rho$.  While the two parametrizations are indistinguisable for 
heavy nuclei, they predict quite distinct patterns for $A=3$.
This is exhibited in Figure 4, which shows the isoscalar EMC
ratios of $^3$H and $^3$He.
The solid curve in Fig. 4 assumes that the EMC effect scales 
with $A$ and describes both $A=3$ nuclei.  The dashed and 
dot-dashed curves assume that the EMC effect scales with $\rho$,
applied to $^3$He and $^3$H, respectively.  Also shown in Fig. 4 
are the recent DESY-Hermes data~\cite{ac00} on the EMC effect for  
$^3$He. 
The expected precision ($\pm1\%$) of this experiment for the
$F_2(^3{\rm H})/F_2(^3{\rm He})$
ratio should easily allow to distinguish 
between the two competing parametrizations.
The new measurements should bring a closure to the EMC effect parametrization
issue and provide crucial input for a more complete explanation of the
origin of the EMC effect. 

\section{Summary} 

We discussed possible DIS
measurements with the $A=3$ mirror nuclei using the 11 GeV
upgraded JLab beam.  The measurements 
can determine in an almost model-independent way the
fundamental $F_2^n/F_2^p$ structure function and $d/u$ quark
distribution ratios at high $x$, and distinguish between predictions
based on perturbative QCD and non-perturbative models.
The precision of the measurements can improve dramatically the
quality of parton distribution parametrizations at high $x$,
which are needed for the interpretation of high energy collider data.
The expected data can also test competing parametrizations of the
nuclear EMC effect and provide valuable constraints on models of its
dynamical origin.\\
 

\end{document}